\documentclass[namedreferences]{solarphysics}

\usepackage[hyperref,optionalrh]{spr-sola-addons} 
\usepackage{graphicx}        
\usepackage{color}           

\chardef\us=`\_

\begin{document}

\begin{article}
\begin{opening}

\title{On the Strength and Duration of Solar Cycle 25: A Novel Quantile-based Superposed Epoch Analysis}

\author[addressref={aff1},email={pete@predsci.com}]{\inits{P.}\fnm{Pete}~\lnm{Riley}\orcid{0000-0002-1859-456X}}

\address[id=aff1]{Predictive Science (PSI), San Diego, California, USA.}

\runningauthor{Riley}
\runningtitle{\textit{Solar Physics} Predicting Solar Cycle 25}

\begin{abstract}
Sunspot number (SSN) is an important - albeit nuanced - parameter that can be used as an indirect measure of solar activity. Predictions of upcoming active intervals, including the peak and timing of solar maximum can have important implications for space weather planning. Forecasts for the strength of solar cycle 25 have varied considerably, from it being very weak, to one of the strongest cycles in recorded history. In this study, we develop a novel quantile-based superposed epoch analysis that can be updated on a monthly basis, and which currently predicts that solar cycle 25 will be a very modest cycle (within the 25th percentile of all numbered cycles), with a monthly-averaged (13-month average) peak of $\sim$ 130 (110) likely occurring in December, 2024. We validate the model by performing retrospective forecasts (hindcasts) for the previous 24 cycles, finding that it out performs the baseline (reference) model (the ``average cycle'') 75\% of the time. 
\end{abstract}
\end{opening}

\section{Introduction}
     \label{S-Introduction} 

Solar activity manifests itself in the sunspot cycle, a roughly 11-year change in the number of observed sunspots on the solar surface. These variations correlate well with other measures of activity, such as coronal mass ejections, solar flares, solar energetic particles (SEPs), as well as radiation. This 11-year modulation represents one half of the 22-year Hale cycle, which follows the complete reversal of the solar magnetic field back to it's original polarity. Although an imprecise measurement of activity, sunspot number (SSN) is perhaps the longest, continuously measured parameter, and, thus, provides a unique glimpse into the long-term variations of the Sun. At least indirectly, it provides a measure of the complexity of the photospheric magnetic field, particularly the number of active regions visible on the solar disk. 

The properties and evolution of the solar cycle have been extensively reviewed by \citet{hathaway15a}. Focusing on variations over the last $\sim 270$ years, or so, we note that the relevant variations include: (1) the Gleissberg cycle -- an 80-90-year variation in the amplitudes of the cycles \citep{gleissberg39a}; (2) the Gnevyshev-Ohl effect -- a two-cycle variation with odd-numbered cycles having a higher peak than the preceding even-numbered cycle \citep{gnevyshev48a}; and (3) a general overall increase from cycle 1 through to the present \citep{wilson88a}. Applying these rules heuristically to cycle 25: (1) The Gleissberg cycle suggests that cycle 23 or 24 occurred at a minimum, thus, the peak for cycle 25 should be modestly larger; (2) the Gnevyshev-Ohl effect suggests that cycle 25 should be modestly larger than the cycle 24 peak; and (3) the secular increase since cycle 1 also suggests that cycle 25 should peak slightly higher than cycle 24. To these heuristics, we can add a further: large swings in peak values do not occur from one cycle to the next. That is, the largest peaks over the last 270 years are built upon the superposition of these three modulations. Thus, it seems reasonable to infer that any prediction should be constrained such that it does not vary substantially from the previous one. 

The International Sunspot number, which is also referred to as the Wolf, Zurich, and relative SSN, is a parameter combining the number of sunspots and groups of sunspots. SSN is computed daily using the following expression:

   \begin{equation}  \label{Eq-H-def}
     R = k (10 g + s), 
   \end{equation}
 where s is the number of individual spots, g is the number of sunspot groups, and k is a factor that varies with location and instrumentation \citep{clette14a}. 

Over the last 10 years, SSN has been revised in several significant, and, in some cases, controversial ways. The details are not directly relevant to the present study, to the extent that our analysis could be applied equally well to any variants. Moreover, we do not anticipate that there would be substantial changes to our results or interpretation. The most widely used dataset is the Sunspot Index and Long-term Solar Observations (SILSO) provided by the Solar Influences Data analysis Center (SIDC), a department of the Royal Observatory of Belgium, and this is the dataset we use in the present study. 

Solar cycle prediction techniques can be broadly separated into two classes: Those attempting to predict an ongoing cycle, and those predicting a future cycle that has not yet begun \citep{hathaway15a}. The former is conceptually easier, and, given the data available to constrain the fit, should always return estimates that are more accurate than the latter. Generally, the techniques applied to predicting an ongoing cycle are empirical in nature, such as regression approaches.  In this study we develop a technique that requires the cycle to have already begun. 

Several reviews have discussed the many approaches of making solar cycle predictions. \citet{pesnell12a}, for example, reviewed and analyzed 75 predictions of the amplitude for solar cycle 24, revisiting the topic in subsequent years to assess the accuracy of these predictions (e.g., \citet{pesnell20a}). In general, precursor models performed better than climatological models. Embedded within a more general review of the solar cycle, \citet{hathaway15a} described the main types of solar cycle prediction models, suggesting that, statistical methods generally performed ``only marginally'' better than a baseline model relying on the properties of the average cycle. Most recently, \citet{petrovay20a} discussed the subset of predictions that were, in principle, related to dynamo theory, and further restricted the subset to those that predicted the amplitude of the next maximum no later than the start of the current cycle. Predictions for solar cycle 25 were summarized recently by \citet{nandy21a}, who concluded that the maximum amplitude for cycle 25 was $136.2 \pm 41.6$. This agreed reasonably well with the consensus prediction from the Solar Cycle 25 Prediction Panel (SC25PP), an international group of scientists, who had earlier published a preliminary forecast of $114.6 \pm 10.0$, anticipated to occur in July 2025. These values, however, are in stark contrast to a prediction by \citet{mcintosh20a}, who argued that cycle 25 would be one of the strongest cycles in recorded history. Their initial estimate of 233 (68\% CIs: 204 - 254) was later updated to a value of 190. More recently, and supporting this position, \citet{prasad22a} used a deep learning technique to suggest that cycle 25 will peak at 172 in August 2023. In contrast, \citet{kumar22a} exploited the Waldmeier effect, that is, the observation that the rise time of a sunspot cycle varies inversely with the cycle amplitude, and a physical link with the buildup of the previous cycle’s polar field after its reversal, to infer that the peak of cycle 25 will only be slightly higher than that of cycle 24. Clearly, as cycle 25 progresses, the forecasts have not (yet) collapsed on a single consensus prediction.

In this study, we develop a simple, but intuitive technique for predicting the likely evolution of the current solar cycle, once it is underway. In this initial study, our primary aim is to present the technique, noting that it reproduces reasonable results, consistent with the majority of previous predictions, and in contrast with the popular but far more extreme prediction of \citet{mcintosh20a} and  \citet{prasad22a}. We do not suggest that the proposed technique performs any better than many of the carefully developed statistical and physics-based models that already exist. Instead, we suggest that it serves as a novel fiduciary for many of these techniques and allows us to make several inferences from it. 

We develop the paper as follows. We begin by summarizing the data used to drive the prediction, then discuss the modelling approach we have developed. After this, we describe how the technique is applied to cycle 25. Next, we assess the technique by performing retrospective forecasts of the previous 24 cycles. Finally, we summarise the implications of these results and suggest future studies that can build upon it.

\section{Methodology} 
      \label{S-Methodology} 
      
\subsection{Data}

In this study, we use the monthly and 13-month running averaged sunspot numbers from the Sunspot Index and Long-term Solar Observations (SILSO) dataset, which includes a careful re-calibration of the dataset \citep{clette20a}. 
Data were downloaded at several points during the analysis phase of this investigation, from 01 January 2022 to 10 October 2022. The additional nine months of data that were subsequently added did not alter the results or conclusions in any material way. We also note that the most recent data remain preliminary. This includes the last five data points in the monthly averaged dataset and the last 12 data points in the 13-month running averaged dataset. 

\subsection{Model}

We develop a simple, empirical model for predicting the basic macroscopic properties of the solar cycle. It is only applicable once the solar cycle has started, and its accuracy likely increases with each successive month of data that is added. The model posits that once underway, SSN will evolve in a way such that it maintains a constant quantile position with respect to previous cycles. Thus, a basic requirement is that the data already available lie on a path of constant quantile up the point of the last available data. The model assumes that the constancy of quantile holds when cycles superposed upon one another with solar activity minimum being the ``zero'' in time. This further assumes that the minima can be accurately inferred from the data.  
For convenience, we call this approach the Quantile-Superposed Epoch Analysis (Q-SEA) model. 
 
\section{Results} 
      \label{S-Results} 

\begin{figure}
\centerline{\includegraphics[width=1.0\textwidth,clip=]{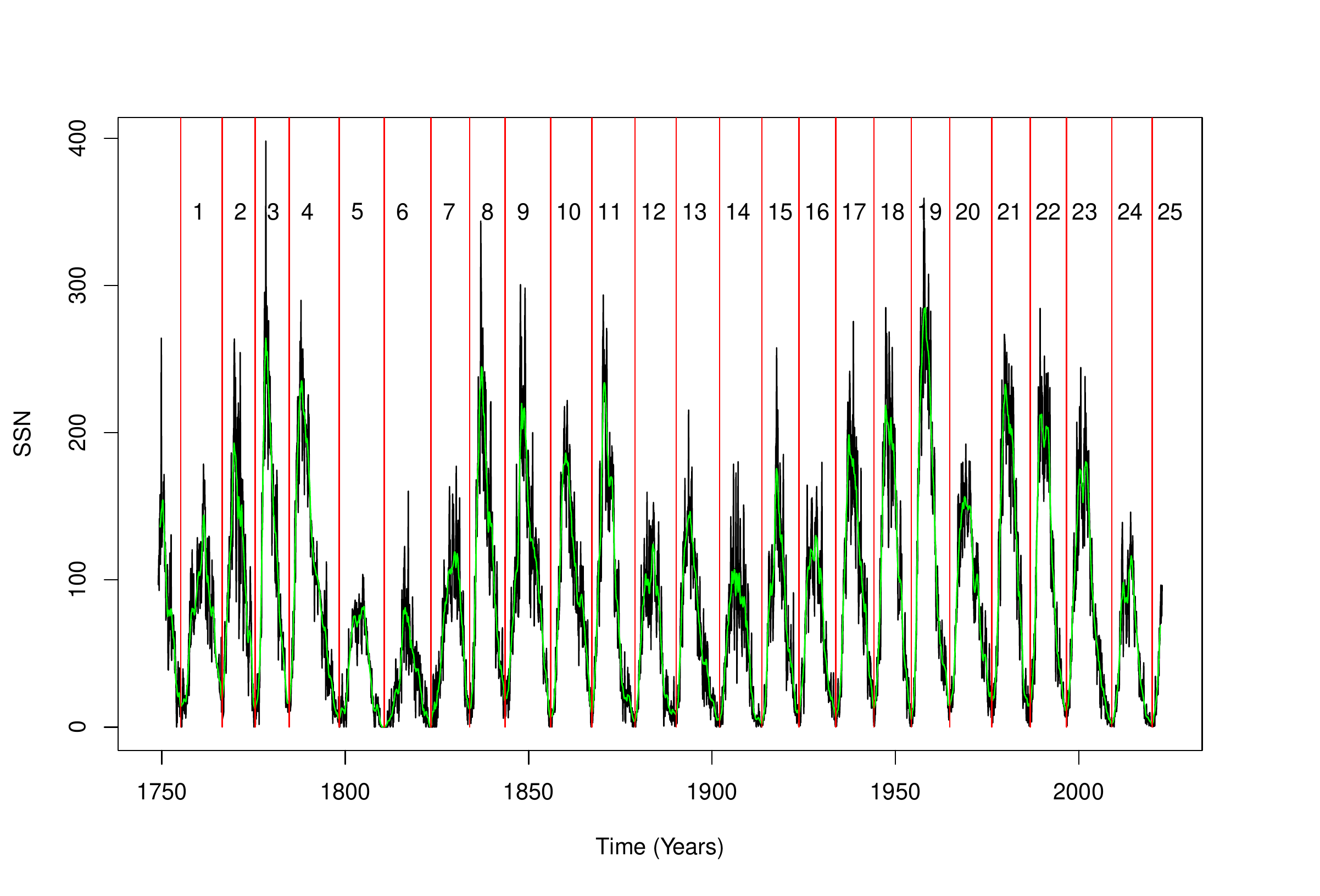}}
\caption{SSN as a function of time over the last $\sim 270$ years. The black curve shows monthly averaged values, while the green curve shows 13-month running averages of the data. The vertical red lines mark the minima in the solar cycles. Solar cycle number is also given.}
\label{fig:ssn-ts}
\end{figure}
      
Figure~\ref{fig:ssn-ts} shows the monthly sunspot number count (black) and 13-month running average (green), with the minima in each solar activity cycle marked by the vertical red line, and cycle numbers indicated between them. The data run through September 2022. As noted above, these are taken from the Sunspot Index and Long-term Solar Observations (SILSO) dataset, which includes careful recalibrations of the data. There are several points worth noting. First, for the last three minima, the depth and duration of the troughs at solar minimum has increased with each minimum, to the point, in fact, that the sunspot number ``bottoms out'' at the beginning of cycle 25 \citep[e.g.,][]{riley22a}.  Additionally, longer-term modulations in the amplitudes of the peaks can be inferred, with cycles 1-6, 6-13, and 13-25 marking three distinct envelopes. Although speculative, these variations suggest that a single small peak is usually followed by several more moderate ones (e.g., cycle 5, 6, and 7), with little evidence that a very small peak (e.g., 24) can be followed by a substantially larger one.

\begin{figure}
\centerline{\includegraphics[width=1.0\textwidth,clip=]{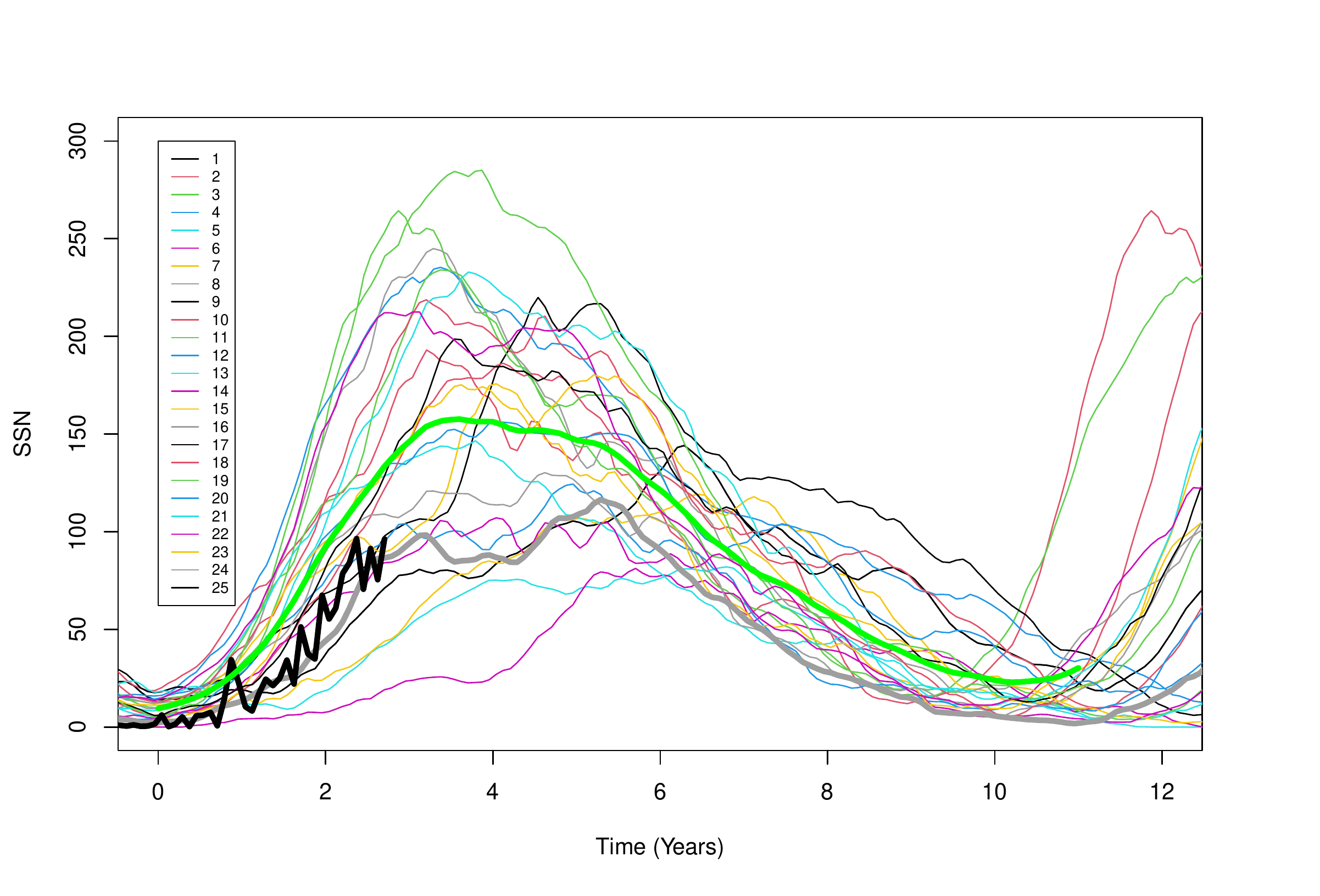}}
\caption{SSN as a function of time for each of the last 25 cycles. The current cycle is presented as monthly averaged, whereas the earlier 24 are shown as 13-month running averages for ease in visualisation. The ``zero'' time for each cycle has been aligned with the estimated minimum in solar activity for that cycle.}
\label{fig:ssn_sm}
\end{figure}

To explore the variation in size and duration in more detail, in Figure~\ref{fig:ssn_sm}, we present a superposed epoch analysis of the last 25 cycles as a function of time from the point of solar minimum. Cycles 1-24 are shown as 13-month averages to aid in their interpretation (we show the monthly-averaged data later), while cycle 25, which is ongoing, is shown as monthly averages (thick black curve). The average of all 24 cycles is shown in green and cycle 24 is emphasized by the thicker grey line. Several points are worth remarking upon. First, there is a huge range of peak values, from below 100 to almost 300. Second, the peaks occur at various points along the cycle, but there is a tendency for larger cycles to peak sooner than weaker cycles. Third, cycles show considerable variability in duration, as evidenced from the location of the following minimum (right-hand-side of the plot), which occurs, roughly between years 9 and 12. Fourth, focusing on cycle 25, we infer that it is increasing along a similar trajectory as cycle 24. Although the less-averaged points do cross multiple previous cycles, in general, there appear to be six curves below it and 18 curves above it. Thus, it lies roughly at the 25th percentile. 
Finally, we note that the current cycle appears to be substantially less active than the average of all 24 previous cycles (green). 

To investigate the quantile position more quantitatively, we interpolated each cycle onto a higher-resolution time-series (varying $\Delta t$ from 0.01 to 0.1 years to verify that the results were not sensitive to the procedure) using the monthly datasets and ranked cycle 25 data at each point in time. Although there was initially some variation due to the generally low values, once the cycle began to take off, cycle 25 consistently placed at rank 6 out of 25. Specifically, for the interval 13-24 months, the modal quantile was found to be overwhelmingly 25\% (6/24).

\begin{figure}
\centerline{\includegraphics[width=1.0\textwidth,clip=]{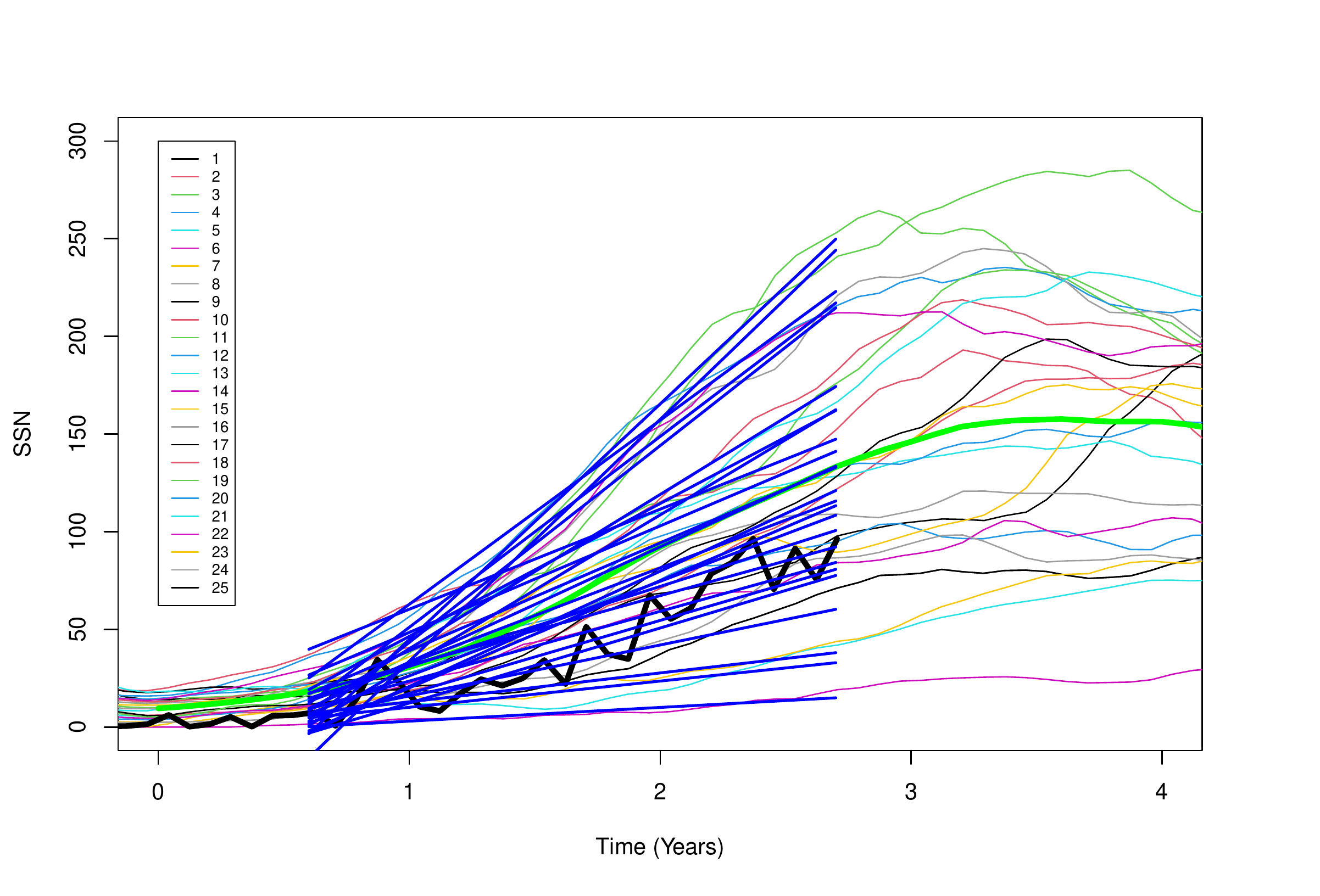}}
\caption{SSN as a function of time for the first four years of each cycle. The solid blue lines are linear fits to each of the curves for the time interval shown.}
\label{fig:ssn-zoom}
\end{figure}

To verify the robustness of this result, we then fit a two-year interval commencing nine months after the beginning of the cycle using either an exponential model or a linear model. The former, unfortunately, did not visually capture a significant fraction of the cycles since the point of exponential rise differed from one cycle to the next. On the other hand, a linear fit seemed to visually perform reasonably well. These results are shown in Figure~\ref{fig:ssn-zoom}, where four years of each cycle are shown, with the fits to these curves shown by the blue lines. Cycle 25, in terms of gradient, ranked 7th out of 25.

\begin{figure}
\centerline{\includegraphics[width=1.0\textwidth,clip=]{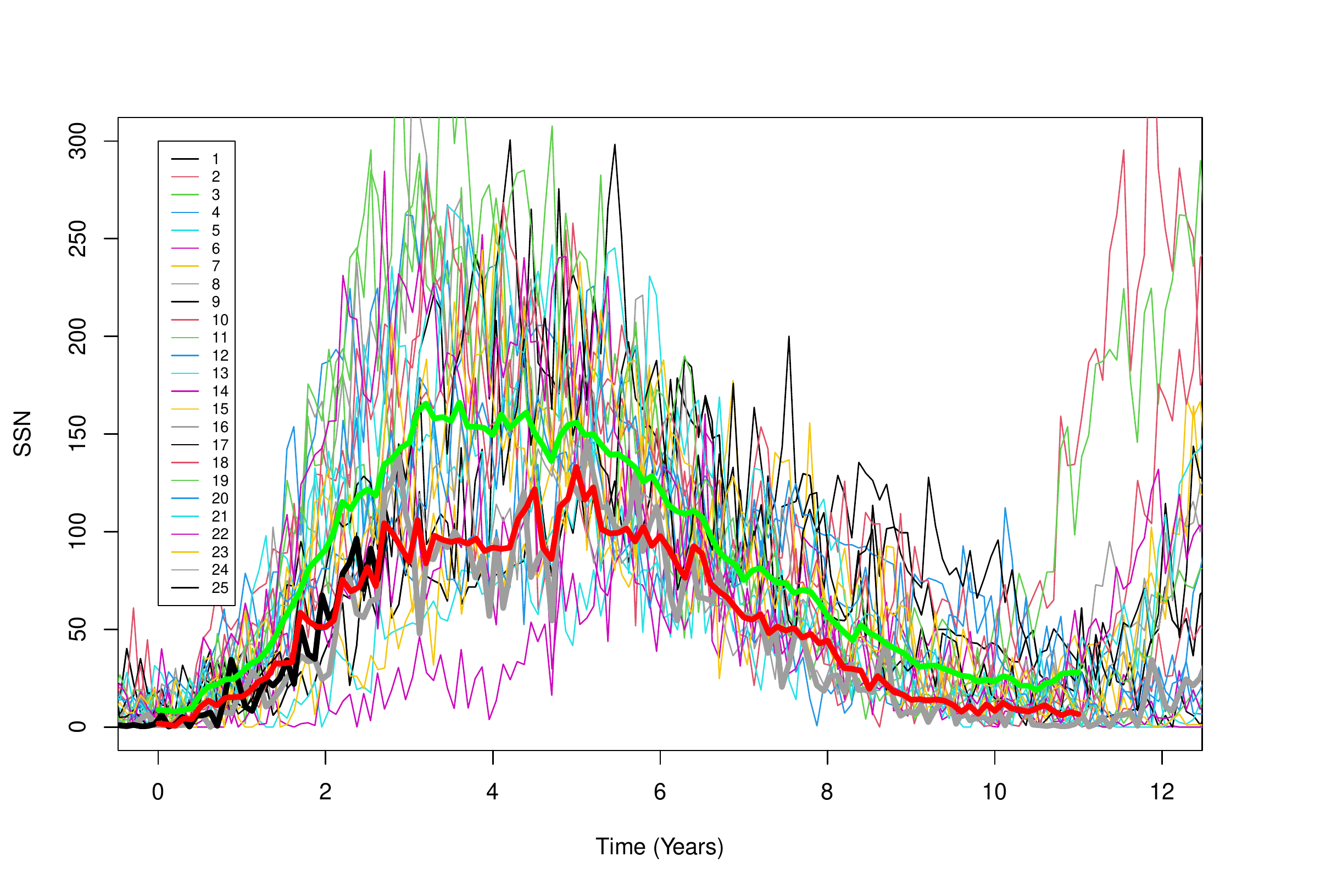}}
\caption{As Figure~\ref{fig:ssn_sm}, but with monthly (not 13-month running averages) shown for each previous cycle. The thick black line again shows observed SSN from cycle 25, while the thick red line shows the predicted Q-SEA values. For comparison, cycle 24 is highlighted by the thicker grey line, and the average cycle is shown in green.}
\label{fig:ssn-qsea}
\end{figure}

Based on these estimates of the quantile position of cycle 25, relative to previous cycles, we then built a model prediction for the entire cycle such that, at each point along the cycle, the SSN would be the 25th percentile of all values at that point in time. This prediction is shown by the red curve in Figure~\ref{fig:ssn-qsea}. In this case, we used the monthly averages for all cycles to make the prediction. Also emphasized in the plot is the variation for cycle 24 (thick grey line), which can be seen to track the predicted trajectory for cycle 25, with a peak value of $\sim 120$. The average cycle (using the previous 24 cycles) is shown by the thick green line. Of particular note, is that the observed values for cycle 25 (thick black curve), which were not used in the prediction, overlay the prediction for cycle 25 remarkably well. Additionally, both the observed and predicted values for cycle 25 lie significantly below the average cycle values.

\begin{figure}
\centerline{\includegraphics[width=1.0\textwidth,clip=]{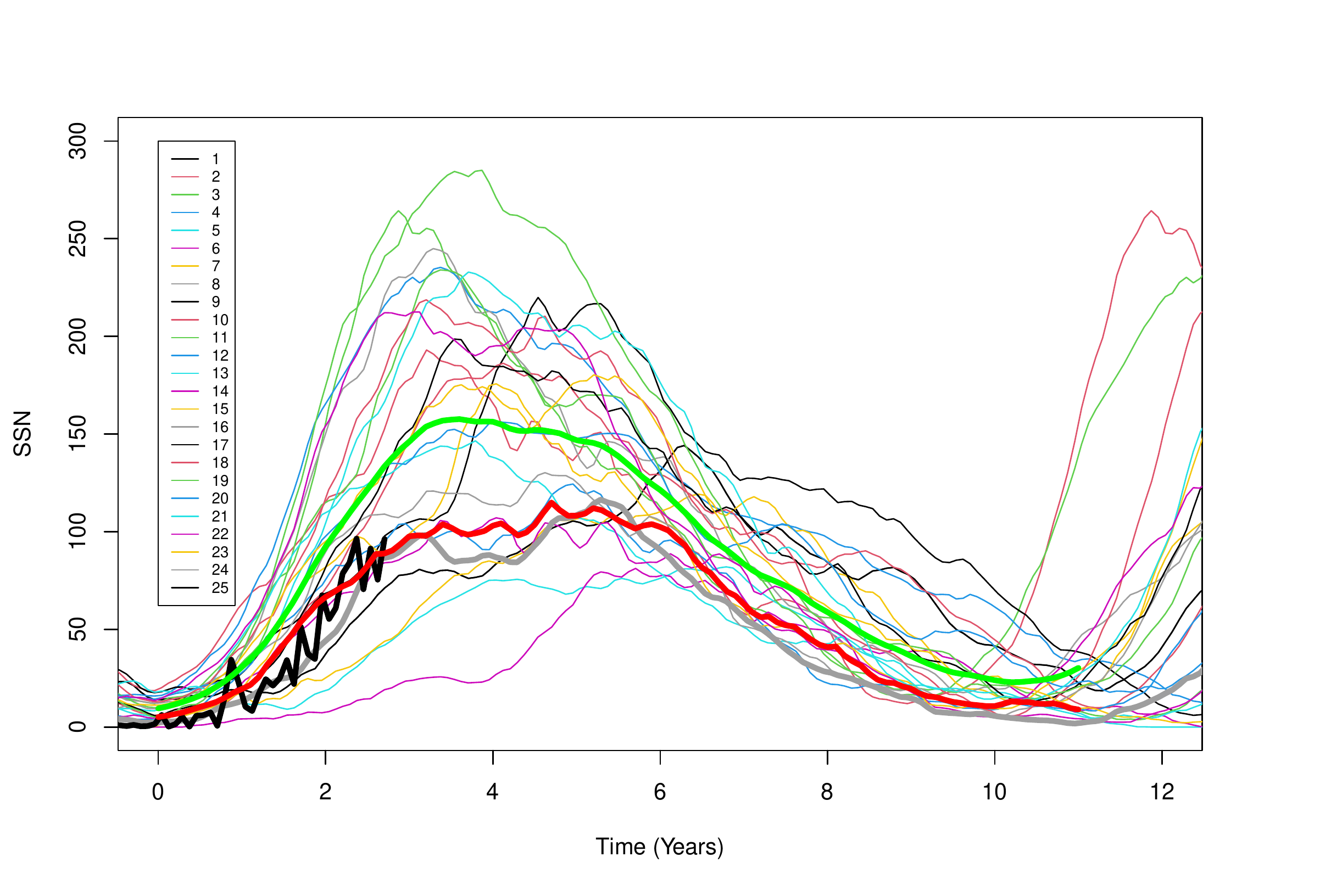}}
\caption{As Figure~\ref{fig:ssn-qsea}, but with 13-month running averages shown for each previous cycle. The thick black line again shows observed monthly SSN from cycle 25, while the thick red line shows the predicted Q-SEA values based on the 13-month data.}
\label{fig:ssn-qsea-13}
\end{figure}

We repeated the analysis using the 13-month running average of the previous cycles. This is shown in Figure~\ref{fig:ssn-qsea-13}. Again, the predicted cycle is relatively weak and tracks the evolution of cycle 24 reasonably well. In this case, comparison between the monthly values for cycle 25 (black curve) and the predicted 13-month running averages (red curve) do not lie on top of each other as well as in Figure~\ref{fig:ssn-qsea}; however, this is likely a statistical artifact of comparing monthly and 13-monthly averages. 

\begin{figure}
\centerline{\includegraphics[width=1.0\textwidth,clip=]{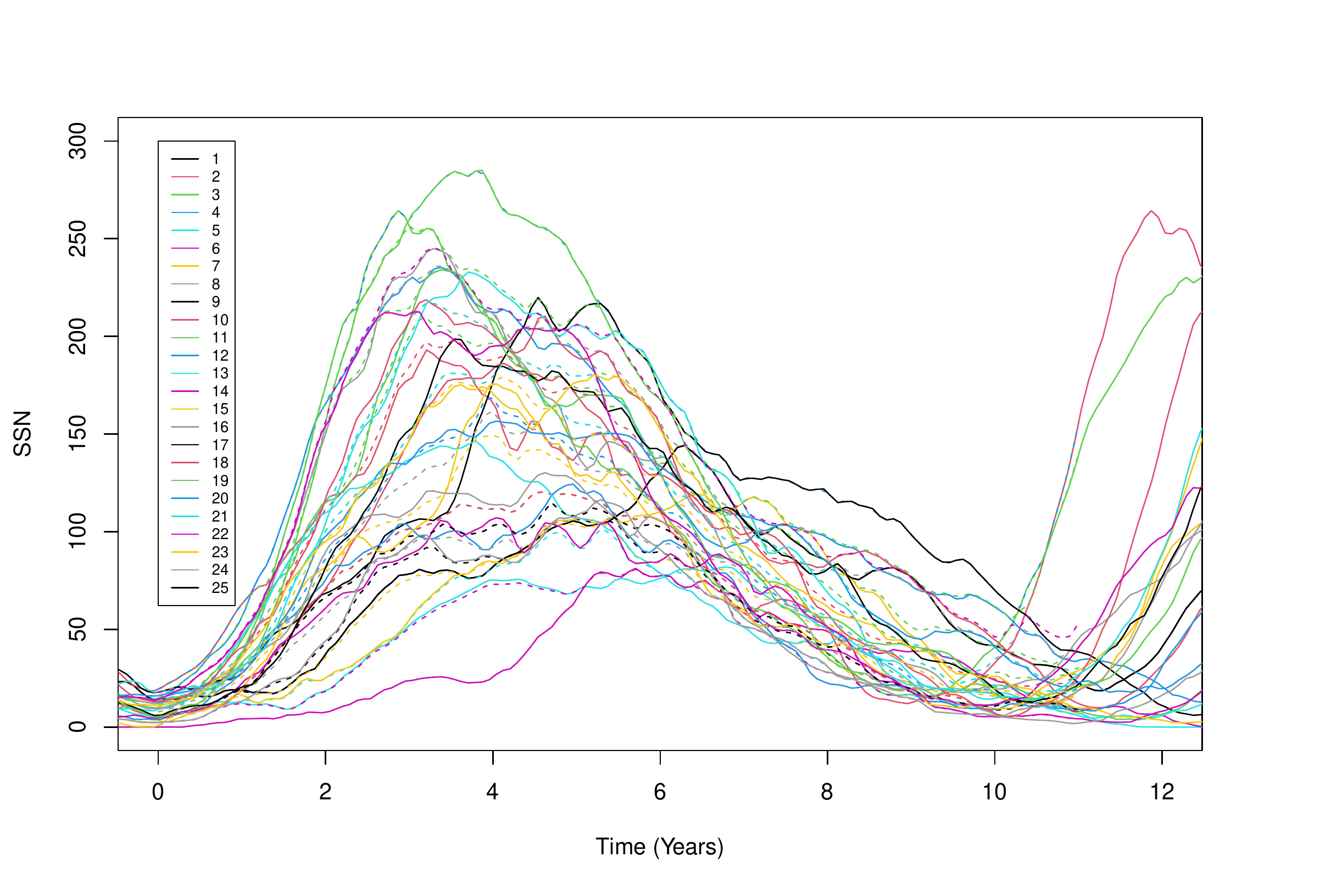}}
\caption{Comparison of 13-month averaged SSN with predicted profile using the Q-SEA method. Data are shown by the solid lines while dashed lines represent the forecasts. The thick green line again shows the }
\label{fig:valid1}
\end{figure}

While suggestive, the results so far merely imply a promising technique. To better assess and validate the model, we repeated the analysis for each of the previous 24 cycles, assuming that only the first two years of the cycle had elapsed. To estimate the quantile for each cycle, we took the modal quantile value for months 13-24. In all cases, there was a clear modal value that could be used to build up a forecast for that cycle. Figure~\ref{fig:valid1} compares the observed SSN profile (solid line) with the predicted profile (dashed line), where each colour denotes a particular cycle. In general, at least in a qualitative sense, there appears to be a reasonable match between predictions and observations. 

\begin{figure}
\centerline{\includegraphics[width=1.0\textwidth,clip=]{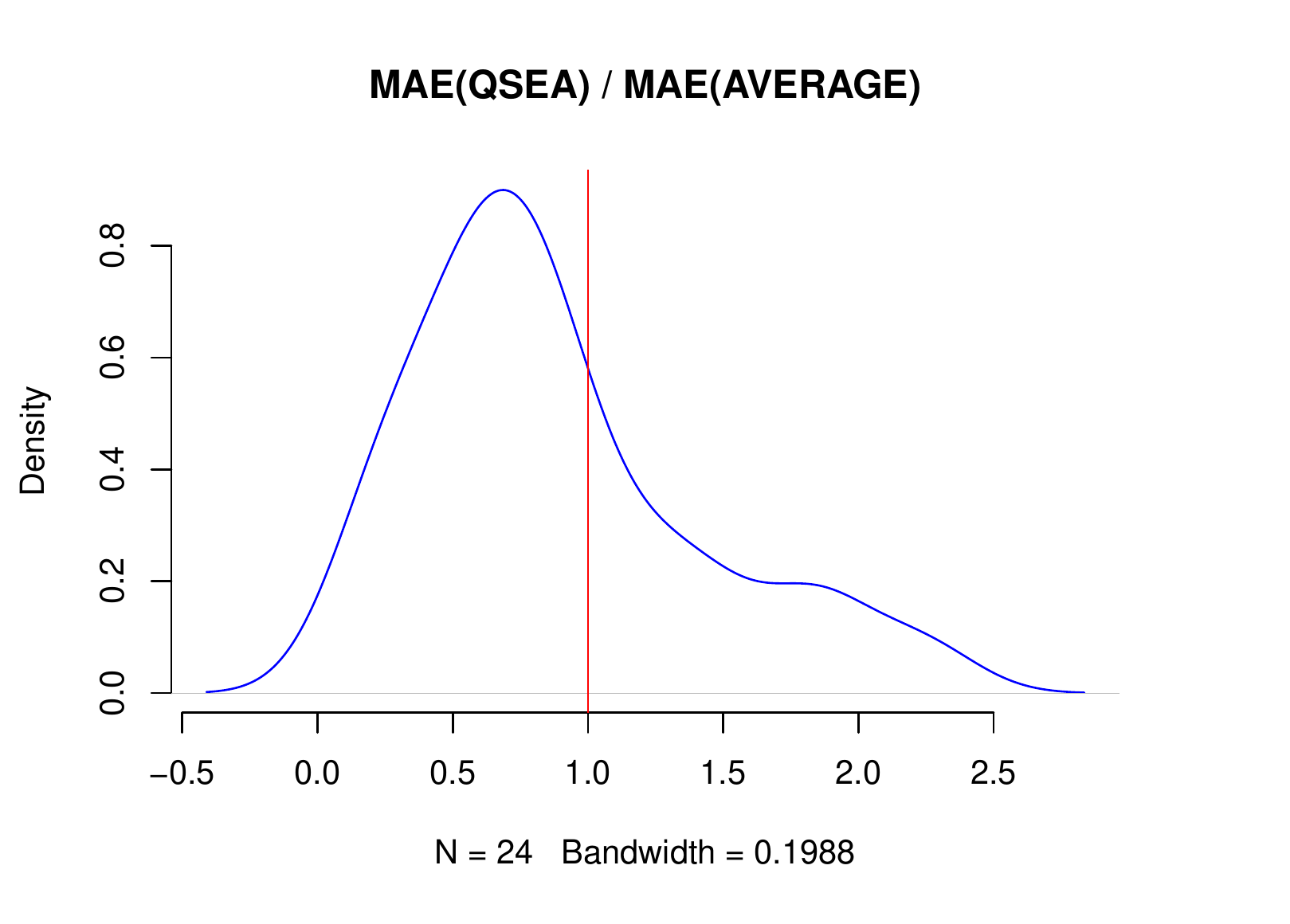}}
\caption{Density distribution of the ratio of MAE(QSEA) versus MAE(Average), where the latter refers to a prediction based on the average value of SSN for all 24 cycles.}
\label{fig:valid2}
\end{figure}

To quantify this, we than computed the mean absolute error (MAE) between the forecasted SSN profile and the SSN profile that was observed. We also computed the MAE for the ``baseline'' (or ``reference'') model, which we choose to be the average cycle profile (the green curves in previous Figures).  The degree to which the former errors are less than the latter provides support for the Q-SEA model. 
In Figure~\ref{fig:valid2} we show the ratio of MAE(Q-SEA) to MAE(average). A distribution centred on 1.0 would suggest that the Q-SEA model was no better than the baseline model; however, a distribution shifted to the left, would favour Q-SEA. We found that 75\% (18 out of 25) cycles were better predicted using the Q-SEA approach. We note also that the area under the curve is not representative of the relative merits of the two models since a ratio is being displayed (thus, for example, a value of 2.0 for MAE(average) would be equivalent to a value of 0.5 for MAE (Q-SEA)). 

\section{Summary and Discussion} 
      \label{S-Discussion} 

In this study, we developed a novel superposed epoch analysis that can be updated on a monthly basis, and which currently predicts that solar cycle 25 will be a very modest cycle (within the 25th percentile of all numbered cycles), with a monthly-averaged (13-month average) peak of $\sim 130 (110)$, likely occurring in December, 2024.  These results are generally consistent with most predictions that suggest cycle 25 will be a weak cycle, but contradict several prominent and more extreme predictions arguing that the current cycle will rival the strongest cycles in recorded history. 

The simple model explored here has several potentially limiting assumptions. The first, is whether it is reasonable to assume that once underway, a cycle will maintain its relative order with respect to other cycles? The dynamo is  intrinsically stochastic in nature, and, thus, it is not unreasonable to believe that random fluctuations below the surface could result in significant, long-term and large-scale changes to the pattern of active regions. 
However, comparison of the Q-SEA model with a baseline profile derived from the ``average cycle'' further supports the value of the approach, with it outperforming the latter 75\% of the time. This suggests that while deviations can occur, a cycle seems to maintain its quantile position reasonably well. 
A second assumption is that we can accurately identify the ``true'' minimum in the cycle, and, further, that this is a reliable marker for the onset of the next cycle. It is conceivable that a different metric of this would lead to different conclusions. However, the close alignment of the observed data for cycle 25 and the predicted values during the entire 2.5 years since the minimum of December 2019 suggests that this is a reasonable assumption to make. 

Our primary purpose in this study was to present the Q-SEA technique and contrast its predictions with previous attempts. We also presented a retrospective forecast, suggesting that it performs better than a baseline model. Pragmatically, however, as \citet{hathaway15a} has suggested, it is likely that the Q-SEA model will fair only as well as previous statistical approaches have, being limited -- in one form or another -- to extrapolation of patterns that are already present in the historical dataset. Instead, to overcome these intrinsic limitations, it is likely that mechanistic models are required, ones that incorporate the non-linearity of the system, as well as being driven by the widest array of available data, such as observations of the photospheric magnetic field.

Our analysis, interpretation, and conclusions differ substantially from those of \citet{mcintosh20a}, who suggested that ``sunspot Solar Cycle 25 could have a magnitude that rivals the top few since records began''. Instead, our results support a growing chorus of studies that appear to rule out this possibility (e.g., \citet{ahluwalia22a,carrasco22a}). Interestingly, even now, the Mcintosh et al. interpretation is highlighted disproportionately in both scientific presentations, news publications, and social media \citep[e.g.][]{phillips22a, pultarova22a}. Its current popularity, while a natural consequence of extreme predictions being more ``newsworthy'', also benefits from a fortuitous alignment of a short section of the current rise portion of the cycle with their prediction. Figure~1 from  \citet{phillips22a}, for example, suggests that, indeed, the \citet{mcintosh20a} forecast  appears to match a limited portion of the rise phase very well. However, a crucial oversight is the mismatch from solar minimum (12/2019) until mid 2021, where their prediction underestimated the observed values.  

In closing, we reiterate that our primary goal was to present a simple technique for predicting the evolution of the solar cycle by placing the current cycle at a particular quantile, and extrapolating from that using the available previous cycles. This approach is, in many ways, complementary to other empirical-based approaches that are applicable only after the cycle has begun. Our aim has not been to demonstrate that this approach is necessarily better than any of these approaches. This will be the topic of a future investigation. Instead, we present it as a logical extrapolation procedure, which is consistent with many of the predictions that have been previously made, including the consensus prediction (subject to a shift in timing) produced by NOAA. It is also presented as a counter to more extreme predictions suggesting that the current cycle is going to rival the largest cycles of the space era. Our results suggest that this, while possible, is extremely unlikely. 

 

\begin{fundinginformation}
The author gratefully acknowledges support from NASA (80NSSC18K0100, NNX16AG86G, 80NSSC18K1129, 80NSSC18K0101, 80NSSC20K1285, 80NSSC18K1201, and NNN06AA01C), NOAA (NA18NWS4680081), and the U.S. Air Force (FA9550-15-C-0001).
\end{fundinginformation}

\begin{dataavailability}
All data analysed in this study can be downloaded from \url{https://www.sidc.be/silso/datafiles}. 
However, for convenience, and to the extent that these data may change, we are also making the data available with the source code through a GitHub repository. 
\end{dataavailability}


\begin{codeavailability}
All code necessary to reproduce the figures and analysis presented here is being made available through a GitHub repository. 
\end{codeavailability}

\begin{ethics}
\begin{conflict}
The authors declare that they have no conflicts of interest. 
\end{conflict}
\end{ethics}


\bibliographystyle{spr-mp-sola}

\bibliography{/Users/pete/Dropbox/manuscripts/references/riley-refs-v3-3.bib} 

\end{article} 

\end{document}